\documentclass{article}
\usepackage{amsmath}
\usepackage{amssymb}
\usepackage{amsfonts}

\providecommand{\U}[1]{\protect\rule{.1in}{.1in}}

\input{tcilatex}

\begin{document}

\title{Variation of the fine structure constant caused by expansion of the
Universe.}
\author{Ivan A. Cardenas, Anton A. Lipovka \\
%EndAName
Department of Investigation for Physics, \\
Sonora University, Sonora, Mexico. \\
e-mail: aal@cifus.uson.mx}
\maketitle

\begin{abstract}
In present paper we evaluate the fine structure constant variation, that
should take place as the pseudo-Riemannian universe expands and its
curvature is changed adiabatically. Such variation of the fine structure
constant is attributed to an energy losses by an extended physical system
(consist of baryonic component and electromagnetic field) due to expansion
of our Universe. Obtained ratio $\overset{\cdot }{\alpha }/\alpha =1\cdot
10^{-18}$ (per second) is only five times smaller than actually reported
experimental limit on this value. For this reason obtained variation can
probably be measured within a couple of years. To argue the correctness of
our approach we calculate the Planck constant as adiabatic invariant of the
electromagnetic field propagated on the pseudo-Riemannian manifold
characterized by slowly varied geometry. Finally, we discuss the double
clock experiment based on $Al^{+}$ and $Hg^{+}$ clocks carried out by
Rosenband et al. (Science 2008). We show that in this case (when the fine
structure constant is changed adiabatically) the method based on double
clock experiment can not be applied to measure the fine structure constant
variation.
\end{abstract}

%%\maketitle


\affiliation{Department of Investigation for Physics, Sonora University, Sonora,
Mexico. e-mail:  aal@cifus.uson.mx}

%%\affiliation{Department of Investigation for Physics, Sonora University}

\section{Introduction}

The extremely important problem of the fundamental constants variation
attracts great attention of the scientific community for the last decades.
Every year a lot of papers on this subject are published both in theory as
well in measurement methods (see [1,2] and references therein). Such an
interest in the subject is due to the huge importance of the problem of the
fundamental constants variation for understanding foundation of physics.
Particular attention is paid for the variation of the fine structure
constant, because it is basic parameter for QED and because the experimental
measurements have reached unprecedented accuracy. It is well known, that the
search for variations under discussion is carried out both in laboratories
[3-7] and by using the cosmological data obtained from observed spectra of
distant quasars [1,2,8,9,10]. Unfortunately up to now, such variations have
not been detected yet, but it is important to note that in the last decade,
the accuracy of laboratory measurements has approached closely to the limit
of variation of fundamental constants, that must take place through the
adiabatic change in the geometry of our Universe. For this reason, the need
for a correct theoretical estimate of the fine structure constant variation
due to adiabatic change in geometry is clearly visible.

In this paper we fill this gap and suggest the calculation of the fine
structure constant variation on time, which must take place due to the
adiabatic changes of scalar curvature provoked by expansion of our universe.
The calculations are carried out first time within the framework of the
Riemannian geometry (we suppose in this paper a standard cosmological model
supplied by Friedmann--Lemaitre-Robertson--Walker metric), and the method
applied is significantly new, simplified and rather obvious. In order to
confirm the correctness of the obtained result, we calculate by the same
method the adiabatic invariant for free electromagnetic field (propagating
on the Riemannian manifold characterized by adiabatically changed curvature)
which actually is the Planck constant. As it was mentioned, all calculations
are carried out in the framework of pseudo - Riemannian geometry in which
there is no natural way of introducing the cosmological constant. For this
reason obtained value is differ slightly (by factor 3/2) of their real value
calculated for the Finslerian manifold [11], which contain the cosmological
constant a natural part of geometry. This result argue clearly that the
metric of the world we live in, is not the Riemannian, but Finslerian one.
Finally we explain why this variations are not detected in the experiments
based on comparison of two different frequencies like those discussed in
recent papers [6,7].

\section{Changing of the fine structure constant due to expansion of the
Universe}

Let us consider a system that consists of a classical field located on the
Riemannian manifold characterized by the adiabatically changed curvature. In
this case as it was previously shown [11, 12] (see also the next part of
this paper) such a system is characterized by an adiabatic invariant, which
for the electromagnetic field is actually the Planck constant. Moreover,
this adiabatic invariant depends on the scalar curvature of the Universe
measured in the point of observation and for this reason is varied over time
[11, 12]. The fine structure constant in turn depends on h ($\alpha
=e^{2}/\hbar c$) and for this reason its value also must changes over time.
It should be stressed here, this consideration can be applied not only to
the classical fields (particularly to the electromagnetic field), but also
to any adiabatically isolated system consisting of fields and charged
baryonic matter interacting by means of this field. In this case, parameters
of the system as a whole depend on the manifold variation. So as the
Universe expands, any physical system (for example an atom) will lose its
energy adiabatically due to a very slow change in the curvature of the
manifold.

How large this variation of energy is? To begin, let's make a preliminary
and very rough assessment of the effect of interest to us. Consider a system
which consists of the classical field and characterized by energy $E$
distributed over volume $V$ (we can put $V=1cm^{3}$). In this case the
changing of the energy due to expansion of the Universe is

\begin{equation*}
\frac{\delta E}{E}=-\frac{\delta V}{V}\text{\ \ .}
\end{equation*}

However, for an electromagnetic (EM) wave characterized by momentum\textbf{\ 
}$P$\textbf{\ }one can write:

\begin{equation}
\frac{\delta P}{P}=-\frac{\delta l}{l}\text{ \ \ .}  \tag{1}
\end{equation}

Here in consistence with Hubble relation we take

\begin{equation}
\delta l=Hl\delta t\text{ \ \ .}  \tag{2}
\end{equation}

As is well known, the fine structure constant was introduced by Sommerfeld
as $\alpha =V/c$, where $V$ - is the velocity of electron in the first Bohr
orbit. For this reason, we should consider a hydrogen atom in an expanding
universe. In this case, the momentum taken away from the system (atom) by
the EM field, is taken from the electron, therefore $P=mV=mc\alpha $ , and
one can evaluate

\begin{equation}
\frac{\delta \alpha }{\alpha }\approx \frac{\delta P}{P}\approx -H\delta
t=-2.3\cdot 10^{-18}\delta t\text{ \ \ .}  \tag{3}
\end{equation}

This very simple estimation gives us an idea about the value of variation we
should expect to obtain in general case.

Now let $M$ be an $3$-dimensional $C^{\infty }$ manifold characterized by
scalar curvature $\mathcal{R=}2/R^{2}$, where $R$ is the curvature radius.
(We note here that throughout this paper we are talking about the evolution
of the EM field, characterized by the Faraday tensor of rank=2 which, in
turn, corresponds to the 2D surface. For this reason, we can write $\mathcal{%
R=}2/R^{2}$.)

Let $x$ be a local coordinate on an open subset $U\subset M$ (here $x$ - is
actually the size of the electromagnetic field resonator). Let $T_{p}(M)$
and $T_{p}^{\ast }(M)$ are respectively tangent and cotangent bundles on M,
where $P_{\alpha }\in T_{p}(M)$ and $P^{\alpha }\in T_{p}^{\ast }(M)$ are
covariant and contravariant components of corresponding 4-momentum.

We are interested in variation of the 4-momentum components $P$ as functions
of the Universe radius $R$ and, consequently, of time $t$. By taking into
account relation $x=R\varphi $, ( $R$ is the effective radius of the
universe and $\varphi $ is corresponding small angle in radians),\ we can
write projection of $x$ on tangent and cotangent bundles\textbf{\ }of $M$ as

\begin{equation}
P^{\alpha}=\xi R\sin\varphi  \tag{4}
\end{equation}

\begin{equation}
P_{\alpha }=\xi R\tan \varphi \text{ \ \ ,}  \tag{5}
\end{equation}%
where coefficients $\xi =\frac{2c}{\kappa }$ (here $\kappa =8\pi G/c^{2}$ is
the coupling constant for the Einstein field equations) are written to
comply $\mathcal{R=}\frac{\kappa }{c^{2}}T$ in classical limit, and factor 2
appears from relation $\mathcal{R=}2/R^{2}$. (We note here that equations
(4) and (5) are written for a single nonzero vector component and therefore
are scalar in nature. Here superscript and subscript are intended to
distinguish co- and contravariant components.)

In this case the absolute value of the momentum can be written as

\begin{equation}
P=\sqrt{P_{\alpha }P^{\alpha }}=\frac{2c}{\kappa }R\frac{\sin \varphi }{%
\sqrt{\cos \varphi }}\text{ \ \ ,}  \tag{6}
\end{equation}

where $R=x/\varphi $ is the local (effective) radius of curvature of the
Universe at the point at which the system under consideration is located.

By taking into account that $\varphi\ll1$ for any reasonable laboratory
system, we can restrict our consideration by first and second terms of the
expansion of $\sin\varphi$ and $\sqrt{\cos\varphi}$, then we get

\begin{equation}
P=\xi R\left( \varphi +\frac{\varphi ^{3}}{12}\right) \text{ \ \ .}  \tag{7}
\end{equation}%
\bigskip

As our manifold $M$ expands, the value of $P$ also changes and taking into
account that $x=R\varphi $, we immediately obtain from (7):

\begin{equation}
\delta P=-\frac{c^{3}}{24\pi GR^{3}}\delta R\text{ \ \ .}  \tag{8}
\end{equation}

It should be stressed here - we write this expression for propagating
electromagnetic field localized within a unit volume. Actually this relation
describes the momentum losses by system due to adiabatic changing of the
manifold's curvature.

To evaluate this expression, we need to re-express $R$ through the
observable parameters. Actually we have such a parameter, named as Hubble
constant $H$. But $H$ give us relation for passing trajectory $l$: $\delta
l=Hl\delta t$.

To establish relation between $R$ and $l$ let us imagine a fly walking over
globe with velocity $c$, whereas we inflate the globe such that $\overset{%
\cdot }{R}=c$ too. It is easy to show that in this simple case the
integrated length $l$ is $l=2R$. Actually this is the length which pass a
photon when it propagates on manifold while its curvature is changing due to
expansion.

So in this case our expression can be rewritten as:

\begin{equation}
\delta P=-\frac{cH^{3}}{6\pi G}\delta t\text{ \ \ .}  \tag{9}
\end{equation}

To evaluate variation of the fine structure constant $\alpha $, it should be
noted that historically it was introduced by Sommerfeld as $\alpha =v/c$,
where $v$ is the electron velocity at the first Bohr orbit for the hydrogen
atom. This definition is correct for classical limit $v<<c$ up to 3-rd
digit, and by taking into account the fact that we are interested in the
first digit (actually we calculate the order of magnitude of the variation),
we may accept this definition for our calculation. For this reason the
momentum of electron is

\begin{equation}
P=\frac{m\alpha c}{\sqrt{1-\alpha ^{2}}}\text{ \ \ ,}  \tag{10}
\end{equation}

and varying it we obtain a losses of momentum by electron on the first Bohr
orbit due to adiabatically changing curvature governed by expansion of our
universe (see also [11, 12, 13])

\begin{equation}
\delta P=\frac{mc}{\left( 1-\alpha ^{2}\right) ^{3/2}}\delta \alpha \text{ \
\ .}  \tag{11}
\end{equation}

By substituting this expression into (9), we find

\begin{equation}
\delta\alpha=-\frac{\left( 1-\alpha^{2}\right) ^{3/2}H^{3}}{6\pi Gm}\delta t
\tag{12}
\end{equation}
\bigskip

This is the variation of the fine structure constant on time due to
adiabatically changed curvature of the Riemannian manifold.

It should be stressed here, this expression for $\delta \alpha $ coincide
well with that obtained in $\left[ 12\right] $ (see also $\left[ 11\right] $%
), within the framework of the Einstein-Cartan geometry, if we write it for
the Riemannian manifold (i.e. we should put $\Lambda =0$ for this case).

Namely we have in $\left[ 12\right] $ ($\Lambda =0$):

\begin{equation}
\alpha=\frac{c^{2}}{32\pi^{2}Gm}\mathcal{R}  \tag{13}
\end{equation}

By varying this expression we immediately obtain

\begin{equation}
\delta\alpha=-\frac{H^{3}}{2\pi^{2}Gm}\delta t  \tag{14}
\end{equation}

that perfectly agree with above obtained expression (12).

Direct calculation for $H=73$ $kms^{-1}Mpc^{-1}=2.4\cdot 10^{-18}s^{-1}$
give us value $\overset{\cdot }{\alpha }/\alpha =-1.7\cdot 10^{-18}$ (in 1
second).

This value is about 5 times smaller if compared with reported sensitivity $%
\overset{\cdot}{\alpha}/\alpha<5\cdot10^{-18}$ $[3]$, but the difference is
not so large and we hope the required sensitivity will be achieved within a
couple of years.

\section{Planck constant from the first principles}

Einstein $[14]$ and later Debye $[15]$ at the beginning of XX century have
shown from thermodynamics that electromagnetic field is quantized and this
fact do not depends of the oscillators nature (properties of baryonic
matter). Unfortunately there was not paid duly attention to this result and
historically it was the baryonic component that was quantized first whereas
the electromagnetic field was quantized much later in 1950 by Gupta $[16]$
and Bleuler $[17]$.

As is known, the Planck's constant is the cornerstone of quantum mechanics
that contains and hides the physical meaning of quantization. It was for
this reason that multiple attempts were made to get it from first
principles, but all of them were unsuccessful. We will not describe here the
complete history of the issue, but mention a couple, in our opinion the most
interesting works.

In 1997 Calogero\textbf{\ } $[18]$\textbf{\ }made an attempt to obtain
Planck's constant as a result of the interaction of every particle with the
background gravitation force. However, his assessment was seriously flawed,
because he did not consider the physics of the process. Instead, he used the
phenomenological semi-quantitative approach, confining himself to
expressions derived from dimensional considerations, which cannot be
considered as satisfactory. The Calogero's approach was discussed by Gaeta%
\textbf{\ } $[19]$\textbf{\ }3 years later, but no attempts to improve it
has been made. Unfortunately, this idea has many flaws and encounters many
difficulties when trying to harmonize it with experiments. In their works,
the quantization mechanism was not proposed and it was not explained why and
how exactly the stochastic gravitational interaction affects the
electromagnetically coupled systems (atoms) leading to quantization. Note
here also that the coupling constant of the gravitational interaction is 40
orders of magnitude smaller than it is for electromagnetic one. Therefore,
taking into account the fact that matter in the universe is ionized, it
would make sense first of all to consider the contribution of the stochastic
motion of charged particles through electromagnet interaction. I think such
a statement of the problem would remove all the questions, since in this
case the value of Planck's constant would turn out to be unreasonably large.
This discussion goes far beyond the scope of this article, since it is of
historical interest only. So we back to our calculations.

In this part of paper we show how the electromagnetic field is quantized on
the pseudo - Riemannian manifold, the curvature of which is changed
adiabatically. Namely we obtain from the geometry of our Universe the
adiabatic invariant for Electromagnetic field (which should be identified
with the Planck constant). As it was mentioned in the introduction, the
calculation of the Planck constant value is made with the same method and
for this reason the obtained result serves as an independent verification of
the validity of the applied method.

As we have seen from the first part of this paper, the momentum $P$ and
energy of electromagnetic field propagating on the manifold characterized by
adiabatically changed curvature, are changed on time. This variation
proceeds adiabatically and can be considered as linear function, that is, we
can retain only the first term of the expansion and neglect the corrections
of subsequent orders of smallness.

\begin{equation}
\frac{\delta E}{E}=-\frac{\delta t}{t}  \tag{15}
\end{equation}

From this expression we can immediately write the adiabatic invariant we are
interested in

\begin{equation}
Et=-\frac{\delta E}{\delta t}t^{2}  \tag{16}
\end{equation}

But for free electromagnetic field we have

\begin{equation}
\delta E=c\delta P  \tag{17}
\end{equation}

By substituting $\delta P$ obtained before into this expression we can write
finally for energy in $1$ $cm.^{-3}$

\begin{equation}
Et=\frac{c^{2}H^{3}}{6\pi G}t^{2}=9.93\cdot 10^{-27}\left( erg\cdot s.\right)
\tag{18}
\end{equation}%
for one second in unit volume. It is a very good coincidence with real value 
$h=6.6\cdot 10^{-27}\left( erg\cdot s.\right) $ for such a simple model we
have considered here within the framework of the Riemannian geometry which
is differ of the Finsler geometry by the absence of the cosmological
constant. It should be stressed, we do not include the cosmological constant 
$\Lambda $ into consideration because on the one hand it naturally appears
only in the complete Finsler geometry, on the other hand, this paper is
dedicated mainly to the problem of the fine structure constant variation in
the Riemann geometry, and it is difficult discus here all details of real
geometry of our Universe and nature of cosmological constant. We just note
here that if the actually measured value $\Lambda =1.7\cdot 10^{-56}$ is
taken into account, the obtained here value of the Planck constant will
decrease slightly and reach actually measured value $h=6\cdot 10^{-27}\left(
erg\cdot s.\right) $. The reader can see these details in our previous works
[11, 12].

To conclude this part we stress again that we prove geometrically the fact
that the electromagnetic field is quantized alone even on the expanded
Riemannian manifold. To do this we need not oscillators and baryonic matter.
The only we need for free electromagnetic field to be quantized is
adiabatically changed curvature of manifold.

\section{The $Hg^{+}$ and $Al^{+}$ optical clocks experiment}

In first part of the paper we have shown that the fine -- structure constant
variation due to adiabatically changed curvature of manifold is $\overset{%
\cdot }{\alpha }/\alpha =1.7\cdot 10^{-18}\left( s^{-1}\right) $. As it was
mentioned above, at present time the experimental constrain on the $\overset{%
\cdot }{\alpha }/\alpha $ is very close to calculated value and consist $%
\overset{\cdot }{\alpha }/\alpha <5\cdot 10^{-18}\left( s^{-1}\right) $ $[3]$%
, so, probably within a couple of years experimental facilities will be able
to measure the variation of fine structure constant caused by expansion of
our Universe, discussed above.

However there is another type of experiments based on comparison of
frequencies variation of two optical clocks. Most precise measurements of
this kind were reported by Rosenband et al in 2008 [6] (see also paper [7]
for the same problem) for $Al^{+}$ and $Hg^{+}$ single-ion optical clocks.
In this paper the preliminary constraint on the temporal variation of the
fine-structure constant $\overset{\cdot }{\alpha }/\alpha <5\cdot
10^{-17}\left( yr^{-1}\right) $ were suggested, that actually corresponds to
variation $\overset{\cdot }{\alpha }/\alpha <3\cdot 10^{-25}\left(
s^{-1}\right) $. In this case a reasonable question arises: why variation we
calculate $\overset{\cdot }{\alpha }/\alpha =10^{-18}\left( s^{-1}\right) $
was not measured, whereas (as we have seen before) it inescapably should
appears due to expansion of the Universe? \ The answer on this question is
simple: because the variation proceeds adiabatically. Let us consider this
issue in details by taking as an example the paper [6] (the same way one can
explain the negative result reported in [7]). The authors of paper [6]
reported that they were measuring variation of ratio of frequencies, i.e. $%
\delta \left( \nu _{Al^{+}}/\nu _{Hg^{+}}\right) $.

To make our expressions more clear, let us write $1$ for $Al^{+}$ and $2$
for $Hg^{+}$ . In this case the measured variation can be written as:

\begin{equation}
\delta\left( \frac{\nu_{1}}{\nu_{2}}\right) =\frac{E_{1}}{E_{2}}\left( \frac{%
\delta E_{1}}{E_{1}}-\frac{\delta E_{2}}{E_{2}}\right)  \tag{19}
\end{equation}

where $E_{1}$ and $E_{2}$ are the energies of transitions $i\rightarrow f$
for $Al^{+}$ and $Hg^{+}$ respectively. So

\begin{equation}
\left( \frac{\delta E_{1}}{E_{1}}-\frac{\delta E_{2}}{E_{2}}\right) =\frac{%
\delta\left( E_{1i}-E_{1f}\right) }{E_{1i}-E_{1f}}-\frac {\delta\left(
E_{2i}-E_{2f}\right) }{E_{2i}-E_{2f}}=  \tag{20}
\end{equation}

\begin{equation*}
=-\frac{\delta E_{1i}}{E_{1i}}\frac{E_{1i}}{E_{1f}}\frac{1}{1-\frac{E_{1i}}{%
E_{1f}}}+\frac{\delta E_{1f}}{E_{1f}}\frac{1}{1-\frac{E_{1i}}{E_{1f}}}+\frac{%
\delta E_{2i}}{E_{2i}}\frac{E_{2i}}{E_{2f}}\frac{1}{1-\frac{E_{2i}}{E_{2f}}}-%
\frac{\delta E_{2f}}{E_{2f}}\frac{1}{1-\frac{E_{2i}}{E_{2f}}}\text{ \ \ .}
\end{equation*}

But for adiabatic variation we have $\frac{\delta E_{1i}}{E_{1i}}=\frac{%
\delta E_{1f}}{E_{1f}}=\frac{\delta E_{2i}}{E_{2i}}=\frac{\delta E_{2f}}{%
E_{2f}}$, thus

\begin{equation}
\frac{\delta E_{1i}}{E_{1i}}-\frac{\delta E_{1i}}{E_{1i}}=0  \tag{21}
\end{equation}

and therefore

\begin{equation}
\delta \left( \frac{\nu _{Al^{+}}}{\nu _{Hg^{+}}}\right) =0\text{ \ \ .} 
\tag{22}
\end{equation}

So one can conclude that the geometrical adiabatic variation cannot be
observed in such experiments, when the frequencies of two single-ion optical
clocks are compared. This conclusion is also true for measurements of the
fine structure constant variation by comparing different spectral lines in
the spectra of quasars.

\section{Conclusions}

In present paper we calculate variation of the fine structure constant which
must take place due to expansion of the Universe. For the pseudo --
Riemannian manifold it consist $\overset{\cdot }{\alpha }/\alpha =1.7\cdot
10^{-18}\left( s^{-1}\right) $ that only 5 time smaller than currently
established constrains on this value $\overset{\cdot }{\alpha }/\alpha
<5\cdot 10^{-18}\left( s^{-1}\right) $ [3].

We also show that on the pseudo -- Riemannian manifold there exist adiabatic
invariant for electromagnetic field which depends on the curvature and has a
value very close (it differ by factor 3/2) to the laboratory measured Planck
constant. Exact value for the Planck constant, as function of curvature and
cosmological constant, can be calculated only within the framework of the
complete Finslerian geometry (which includes the cosmological constant in a
natural way) and can be found in [11] and [12] . This suggests that we live
not in the (pseudo-) Riemannian world, but in Finsler one.

It is shown that double clock experiment as well as the attempt to measure
the variation of the fine structure constant by observing the spectra of
quasars (by comparing several spectral lines), cannot be used to measure an
adiabatically changed values (particularly these techniques can not be
applied to measure the fine structure constant variation).


\begin{thebibliography}{10}
\bibitem[1]{1} S. A. Levshakov, arXiv:1603.01262 (2016)

\bibitem[2]{2} P. Bonifacio et al., Astronomische Nachrichten \textbf{335},
83 (2014)

\bibitem[3]{3} D. R. Leibrandt et al., Phys. Rev. Lett. \textbf{111}, 237402
(2013)

\bibitem[4]{4} S. Cook, T. Rosenband, and D.R. Leibrandt, Phys. Rev. Lett. 
\textbf{114}, 253902 (2015)

\bibitem[5]{5} Y. V. Stadnik and V. V. Flambaum, Phys. Rev. A \textbf{93},
063630 (2016)

\bibitem[6]{6} T. Rosenband et al., Science \textbf{28}, 319(5871) (2008)

\bibitem[7]{7} R.M. Godun et al., Phys. Rev. Lett. \textbf{113}, 210801
(2014).

\bibitem[8]{8} P. A. R. Ade et al., Astronomy \& Astrophysics \textbf{580},
A22 (2015)

\bibitem[9]{9} H. Rahmani et al., Monthly Notices of the Royal Astronomical
Society \textbf{425}, 556, DOI: 10.1111/j.1365-2966.2012.21503.x (2012)

\bibitem[10]{10} S.A. Levshakov et al., Astronomy \& Astrophysics \textbf{540%
}, id.L9, (2012) DOI: 10.1051/0004-6361/201219042

\bibitem[11]{11} A.A. Lipovka, Journal of Applied Mathematics and Physics 
\textbf{5}, 582, (2017) doi: 10.4236/jamp.2017.53050. arXiv:1608.04596

\bibitem[12]{12} A.A. Lipovka, Journal of Applied Mathematics and Physics 
\textbf{2}, 61 (2014) doi: 10.4236/jamp.2014.25009.

\bibitem[13]{13} A.A. Lipovka, Journal of Applied Mathematics and Physics 
\textbf{4}, 897 (2016) doi: 10.4236/jamp.2016.45098.

\bibitem[14]{14} A. Einstein, Annalen der Physik \textbf{17}, 132 (1905)

\bibitem[15]{15} P. Debye, ) Annalen der Physik \textbf{33}, 1427 (1910

\bibitem[16]{16} S. Gupta, Proc. Phys. Soc. \textbf{63}A (7), 681, (1950)
Bibcode:1950PPSA...63..681G doi:10.1088/0370-1298/63/7/301

\bibitem[17]{17} K. Bleuler, Helv. Phys. Acta (in German) \textbf{23} (5),
567 (1950), doi:10.5169/seals-112124

\bibitem[18]{18} F. Calogero, Cosmic origin of quantization, Phys. Lett. A 
\textbf{228} 335 (1997)

\bibitem[19]{19} G. Gaeta, International Journal of Theoretical Physics 
\textbf{39}, 1339 (2000)
\end{thebibliography}
\end{document}